\documentclass[reprint]{revtex4-1}

\usepackage{amsmath}    % need for subequations
\usepackage{graphicx}   % need for figures
\usepackage{verbatim}   % useful for program listings
\usepackage{color}      % use if color is used in text
\usepackage{subfigure}  % use for side-by-side figures 
\usepackage{hyperref}   % use for hypertext links, including those to external documents and URLs

\begin{document}

\title{Observation of cyclotron resonance and measurement of the hole mass in optimally-doped La$_{2-x}$Sr$_{x}$CuO$_4$}

\author{K.~W. Post$^1$, A. Legros$^{2}$, D.~G. Rickel$^1$, J. Singleton$^1$, R.~D. McDonald$^1$, Xi He$^{3,4}$, I.~Bo\v{z}ovi\'{c}$^{3,4}$, X. Xu$^{3,5}$, X. Shi$^{5}$, N.~P. Armitage$^{2,6*}$, S.~A. Crooker$^{1*}$}
\affiliation{$^1$National High Magnetic Field Laboratory, Los Alamos National Laboratory, Los Alamos, NM 87545}
\affiliation{$^2$Department of Physics and Astronomy, The Johns Hopkins University, Baltimore, MD 21218}
\affiliation{$^3$Brookhaven National Laboratory, Upton, NY 11973, USA}
\affiliation{$^4$Department of Chemistry, Yale University, New Haven, CT 06520, USA}
\affiliation{$^5$Department of Physics, University of Texas at Dallas, Richardson, TX 75080, USA}
\affiliation{$^6$Canadian Institute for Advanced Research, Toronto, Ontario M5G 1Z8, Canada}

\date{\today}

\begin{abstract}
Using time-domain terahertz spectroscopy in pulsed magnetic fields up to 31~T, we measure the terahertz optical conductivity in an optimally-doped thin film of the high temperature superconducting cuprate La$_{1.84}$Sr$_{0.16}$CuO$_4$. We observe systematic changes in the circularly-polarized complex optical conductivity that are consistent with cyclotron absorption of $p$-type charge carriers characterized by a cyclotron mass of $4.9\pm 0.8$ $m_{\rm e}$, and a scattering rate that increases with magnetic field. These results open the door to studies aimed at characterizing the degree to which electron-electron interactions influence carrier masses in cuprate superconductors.
\end{abstract}

\maketitle

The renormalization of effective electronic masses in materials is a well-established consequence of electron-electron and electron-lattice interactions. However, precisely how this renormalization manifests depends on the particular measurement.  Angle-resolved photoemission, quantum oscillation studies in high magnetic fields (such as Shubnikov-de Haas or de Haas-van Alphen measurements), and heat capacity all measure masses that reflect the underlying renormalized quasiparticle dispersion \cite{PinesNozieres}. In contrast, susceptibility or compressibility measurements are sensitive to electronic correlations, but not electron-phonon renormalizations \cite{PinesNozieres}.  In this regard, cyclotron resonance (CR) experiments merit special consideration.  Famously, Kohn showed that in Galilean-invariant systems, CR reveals a cyclotron mass $m_c = eB/\omega_c$ (where $B$ is the magnetic field and $\omega_c$ is the cyclotron frequency) that is \textit{not} affected by electron-electron interactions \cite{kohnCyclotronResonanceHaasvan1961}.  Approximate Galilean-invariance is realized in very low charge density systems (\textit{e.g.}, lightly-doped semiconductors) where the Fermi wavelength greatly exceeds the lattice constant.  Here the band mass plays the role of the free electron mass.  However, these considerations do not apply in higher density systems (\textit{e.g.}, most metals) where, for example, umklapp scattering can be significant \cite{KanakiYamada}.  Moreover, finite disorder or even small deviations from parabolicity can cause electron-electron interactions to manifest in the cyclotron mass \cite{Macdonald}.  The key point is that electron-electron correlations can influence effective masses measured by CR very differently in comparison to other experimental methods.

In the widely studied high-temperature superconducting cuprate (HTSC) materials, numerous transport, spectroscopic, and thermodynamic-based studies of effective carrier masses have been reported \cite{doiron-leyraudQuantumOscillationsFermi2007, vignolleQuantumOscillationsOverdoped2008, barisicUniversalQuantumOscillations2013, sebastianQuantumOscillationsHoleDoped2015, ramshawAngleDependenceQuantum2011, ramshawQuasiparticleMassEnhancement2015, Helm2009, Higgins2018, Loram1994, Michon2019, Girod2020, Ino2002}.  Generally, inferred masses are heavy (of order $1-10~m_e$, where $m_e$ is the bare electron mass), and tend to increase near optimal doping, likely due to renormalization by underlying electron-electron interactions.  As such, CR studies can complement existing methods and help to disentangle the important role of electronic correlations in HTSCs. In the non-interacting limit the cyclotron mass $m_c$ is simply expressed \cite{AshcroftMermin} as $m_c = \frac{\hbar^2}{2\pi} \frac{\partial A}{\partial E}$, where $A$ is the cross-sectional area of the Fermi surface in the plane normal to $B$.  It is also noteworthy that CR places much less stringent requirements on sample quality as compared to most other experimental approaches. Small cyclotron shifts of the frequency-dependent conductivity can be resolved (particularly with the THz experiments described here) even when $\omega_c \tau_{t} $ is significantly less than unity ($\tau_t$ is the transport scattering time). This is in contrast to quantum oscillation studies that -- due to the exponential sensitivity of quantum oscillations to scattering -- generally require $\omega_c \tau_{q} $ approaching unity (where $\tau_q$ is the quantum or phase scattering time \cite{ramshawQuasiparticleMassEnhancement2015}).  Moreover, because it is sensitive to all scatterings and not just current-degrading ones, $\tau_{q}$ is typically smaller than $\tau_t$ \cite{Kartsovnik2004, singletonBandTheoryElectronic2001}. 

However, direct detection of CR in the cuprates has proven challenging. This is because carrier masses are expected to be heavy enough that $\omega_c$ is expected to be small ($\lesssim 30$~GHz/T).  Additionally, scattering times even in high-quality HTSC samples are short, typically $\lesssim$ 1~ps \cite{mahmoodLocatingMissingSuperconducting2019a, Corson2000, Valla2000, Hussey2003, Ramshaw2017}, so that the low-frequency Drude conductivity peak has a broad linewidth $\gtrsim$ 1~THz. Based on these two considerations it is clear that high magnetic fields (many tens of teslas) are needed to make $\omega_{\rm c}$ sufficiently large so that a cyclotron shift of the broad Drude peak can be experimentally resolved.  

To this end, we have combined pulsed magnetic fields with time-domain THz spectroscopy (TDTS) to reveal CR of holes in the normal state of a high-quality thin film of optimally-doped La$_{1.84}$Sr$_{0.16}$CuO$_4$.  The field dependence of the circularly-polarized complex optical conductivity is consistent with CR and a hole mass $m_{\rm c}=4.9 \pm 0.8~m_{\rm e}$. These measurements represent the first direct detection of CR in the La$_{2-x}$Sr$_{x}$CuO$_4$ cuprate superconductors.  Quantum oscillation studies of hole-doped cuprates have generally been limited to low temperatures, while still demanding extremely high magnetic fields and exceptionally clean samples of YBa$_2$Cu$_3$O$_{7-\delta}$ and YBa$_2$Cu$_4$O$_8$ (YBCO) \cite{doiron-leyraudQuantumOscillationsFermi2007, sebastianQuantumOscillationsHoleDoped2015}, Tl$_2$Ba$_2$CuO$_{6+\delta}$, and HgBa$_2$CuO$_{4+\delta}$ \cite{barisicUniversalQuantumOscillations2013}, each of which is limited in its respective doping range.  The ability to perform CR in pulsed magnetic fields therefore enables studies of more disordered samples, and at higher temperatures. This is especially promising in the case of La$_{2-x}$Sr$_{x}$CuO$_4$, which can be doped across the entire phase diagram \cite{Ino2002, Ando2004, bilbroTemporalCorrelationsSuperconductivity2011a}. Prior reports of broadband magneto-optics in HTSC materials were performed at low temperatures and low $B$, and no clear spectroscopic signatures of CR were obtained \cite{spielmanObservationQuasiparticleHall1994, drewFarinfraredMagnetoopticsHighTc1994}. Our terahertz measurements demonstrate the capability to measure CR in La$_{2-x}$Sr$_{x}$CuO$_4$ at temperatures near $T_{\rm c}$ at optimal doping, which opens the possibility of characterizing $m_{\rm c}$ across the entire phase diagram of a single superconducting compound.

We studied a 53~nm thick La$_{1.84}$Sr$_{0.16}$CuO$_4$ film (hereafter termed LSCO), deposited by atomic layer-by-layer molecular beam epitaxy on a 1~mm thick (001)-oriented LaSrAlO$_4$ substrate. The sample was close to optimal doping, and mutual inductance measurements demonstrated a superconducting transition temperature of 41~K.  At this doping level, Hall  measurements show a hole-like response \cite{Ando2004, Balakirev2009}.  Figure 1a shows the experiment, in which a small pulsed magnet is incorporated into the free-space beam path of a TDTS spectrometer.  The field is applied along the film's (001) direction, which is the same as the light propagation direction (\textit{i.e.}, the Faraday geometry).  The magnet has a 15~mm bore and consists of 144 windings of high-strength CuAg wire, and is powered by a purpose-built 20~kJ capacitor bank. The field profile of a 31~T pulse is shown in Fig. 1b. To rapidly detect the THz waveform at peak field, we used a TDTS system based on electronically-controlled optical sampling (ECOPS) \cite{dietzAllFibercoupledTHzTDS2014, Noe2014}, wherein the timing delay between the two ultrafast optical pulse trains that drive the THz emitter and receiver can be electronically modulated very quickly (see Fig. 1c). When synchronized to the magnetic field, the THz waveform can be recorded at multiple field values during a single magnet pulse. Following standard procedure, the measured THz electric field was Fourier-transformed to yield the frequency-dependent complex optical transmission $T(\omega, B)$, which is normalized to the zero-field transmission $T(\omega, 0)$. Signal to noise was further improved by averaging $\approx$20 magnet pulses for each displayed spectrum.

\begin{figure}[tbp]
\centering
\includegraphics[width=0.85\columnwidth]{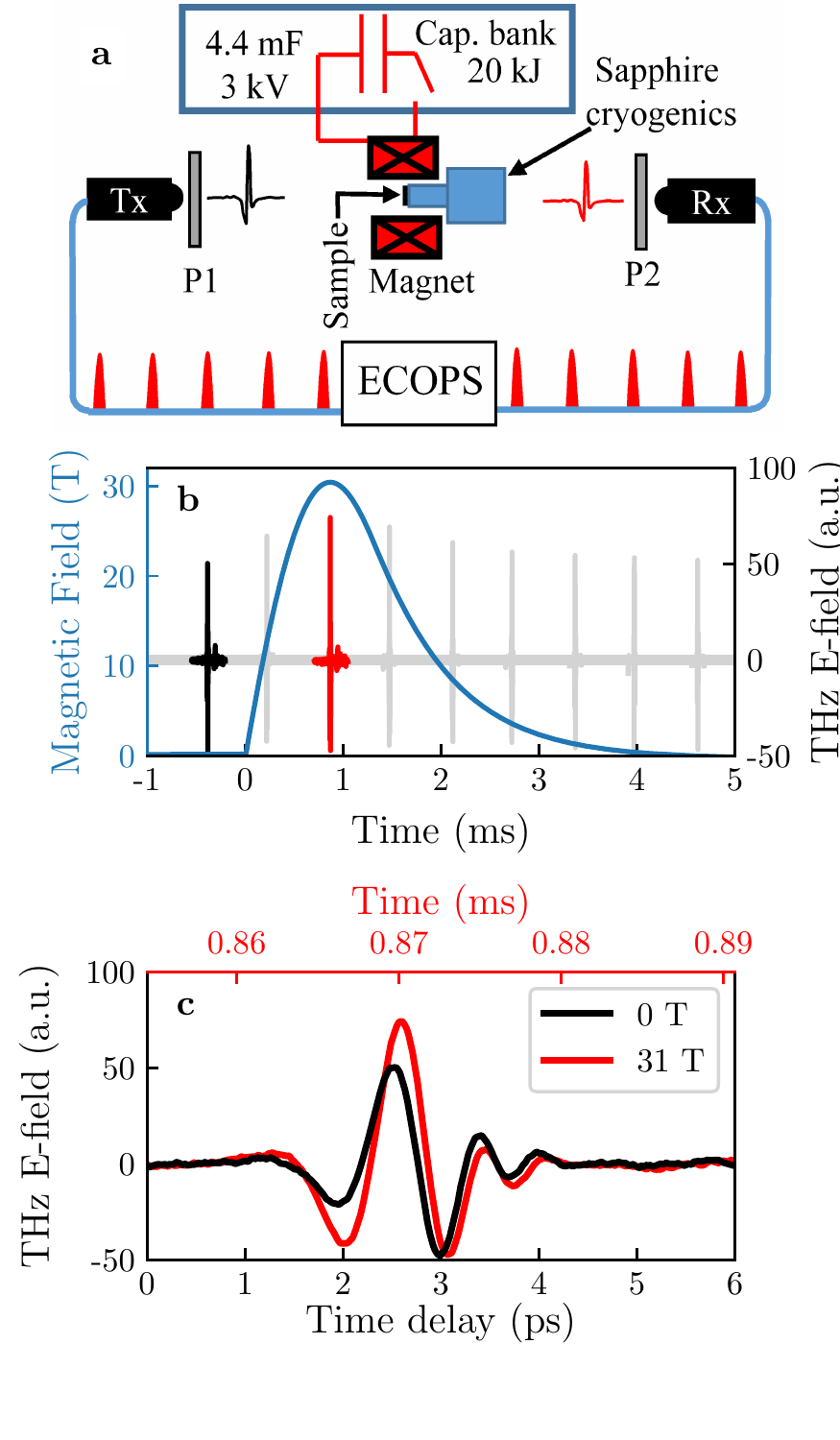}
\caption{\label{Fig1} a) Experimental schematic. A small pulsed magnet is incorporated into the beam path of an ECOPS-based time-domain terahertz spectrometer. Broadband THz optical pulses generated by the transmitter (T$_{\rm x}$) are linearly polarized (by P1) and focused through the sample in the magnet bore, and then directed through a second polarizer (P2) and detected by gated receiver (R$_{\rm x}$).  Parabolic collimating and focusing mirrors are not shown. b) Field profile of a 31~T magnet pulse (blue, left axis) is plotted along with the simultaneously-measured THz electric field (right axis). c) THz electric field measured at zero field (black) and at 31~T (red). The upper x-axis shows actual laboratory time.  The ECOPS system sweeps the effective time delay through 5~ps in $\approx$30~$\mu$s.}
\end{figure}

Crucial to this experiment, we measure the complex conductivity of the LSCO film in the right- and left-circularly polarized bases, which are the natural eigenstates of cyclotron motion [\textit{i.e.}, $\sigma_r (\omega)$ and $\sigma_l (\omega)$, the cyclotron-active and -inactive modes, respectively]. As shown below this allows us to unambiguously resolve small cyclotron frequency shifts of the optical conductivity even in the regime where $\omega_c \tau_t <1$.  This is not possible using conventional infrared spectroscopy that does not measure complex transmission coefficients. Here we obtain $\sigma_r (\omega)$ and $\sigma_l (\omega)$ by measuring both diagonal and off-diagonal complex transmission components $T_{xx}(\omega)$ and $T_{xy}(\omega)$. Access to $T_{xy}$ is achieved by incorporating two linear polarizers into the THz beam path (P1 and P2). When P2 is rotated $\pm 45^{\circ}$ with respect to P1, then $T_{\pm 45^{\circ}} = (T_{xx}\pm T_{xy})/\sqrt{2}$, from which we can extract $T_{xy}$ and $T_{xx}$, and reconstruct the complex conductivity $\sigma_{r,l} (\omega)$ from the right- and left-circular complex transmission $T_{r,l} = T_{xx} \pm  iT_{xy}$\ as described in Ref. \cite{chengMagnetoterahertzResponseFaraday2019}.   Note that the ability to reconstruct the complex $T_{r,l} $ in a broadband fashion is a capability unique to TDTS and is, as shown below, essential to extracting $\omega_c$ in this material.

\begin{figure}[tbp]
\centering
\includegraphics[width=1.05\columnwidth]{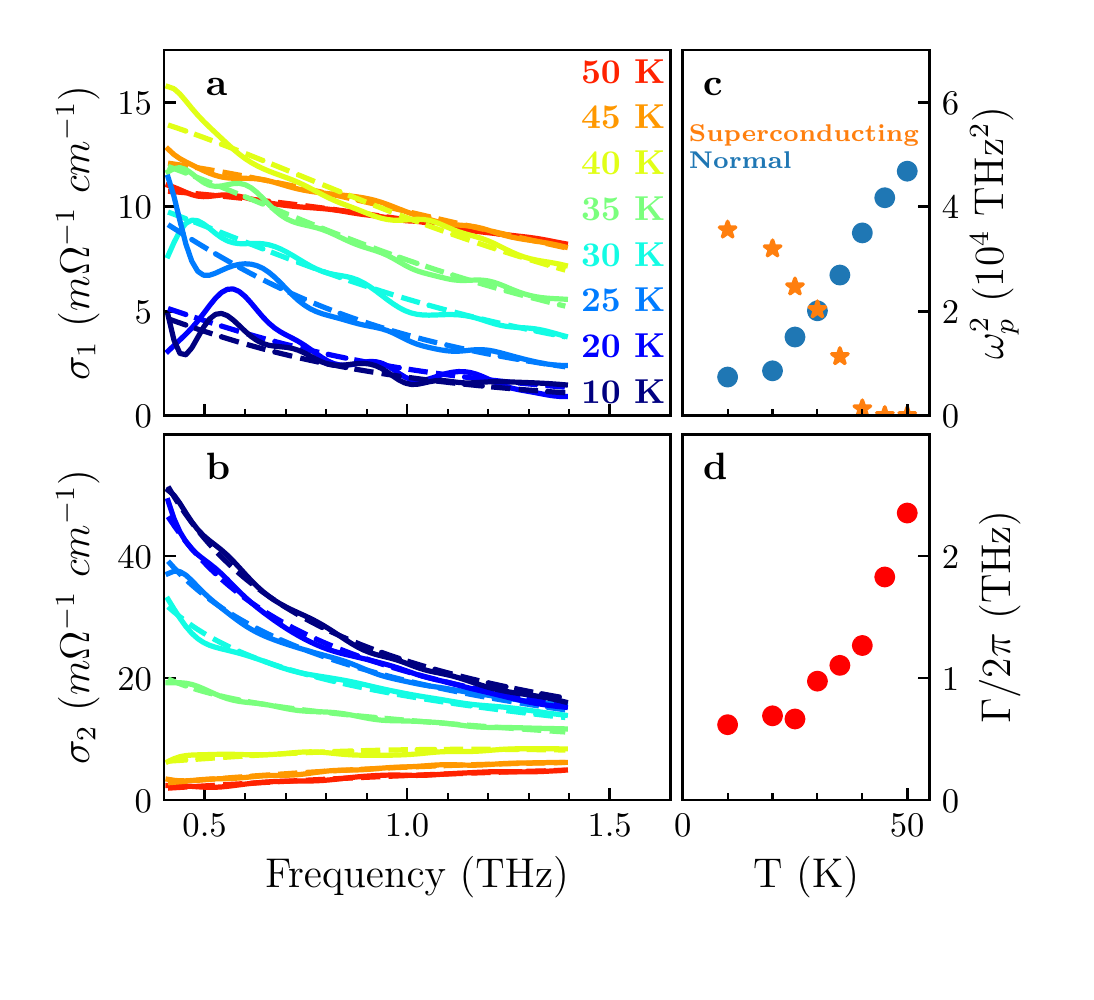}
\caption{\label{Fig2} a,b) The real and imaginary optical conductivity ($\sigma_1$ and $\sigma_2$ respectively) of the LSCO $x$=0.16 film at zero magnetic field, at different temperatures. Dashed lines show fits to the spectra following Eq. 1.  c) Extracted  spectral weight of the normal and superconducting carriers. d) Extracted scattering rate $\Gamma$ of the normal carriers.}
\end{figure}

Figures 2a and 2b show the real and imaginary parts of the complex conductivity ($\sigma_1$ and $\sigma_2$, respectively) of the LSCO sample at different temperatures at zero field. Upon cooling from 50~K to 40~K, there is a sharpening of the Drude conductivity peak in $\sigma_1(\omega)$, typical of a metal with a decreasing scattering rate $\Gamma$. With further cooling to 5~K, the Drude peak loses spectral weight while $\sigma_2(\omega)$ exhibits a $1/\omega$ dependence that increases in magnitude. These are well-established hallmarks of superconductivity, and can be described by a two-fluid model applied to the conductivity \cite{tinkhammichaelIntroductionSuperconductivity2004}:
\begin{equation}
\label{Eqn3}
\sigma(\omega) = i\epsilon_0  \left( \frac{\omega^2_{\rm p,n}}{\omega+i \Gamma} + \frac{\omega_{\rm p,s}^2}{\omega} \right). 
\end{equation}
Here, $\omega^2_{\rm p,n}$ and $\omega^2_{\rm p,s}$ are the spectral weights ($\propto$ plasma frequency squared) associated with the normal and superconducting carrier densities $n_{\rm n}$ and $n_{\rm s}$, and $\Gamma$ is the scattering rate of the normal carriers. Fits are indicated by the dashed lines, and the extracted spectral weights and $\Gamma$ are shown in Figs. \ref{Fig2}c and 2d.  The simultaneous reduction of $\omega_{\rm p,n}^2$ and increase of $\omega_{\rm p,s}^2$ with decreasing temperature reveals the condensation of normal carriers into superconducting Cooper pairs. The point at which $\omega_{\rm p,s}^2$ vanishes corresponds to $T_{\rm c}$ ($\approx 40$~K). The small values of $\Gamma$ (even up to $T_c$) are consistent with those obtained in similar high-quality thin films \cite{mahmoodLocatingMissingSuperconducting2019a}. 

\begin{figure}[t]
\centering
\includegraphics[width=0.8\columnwidth]{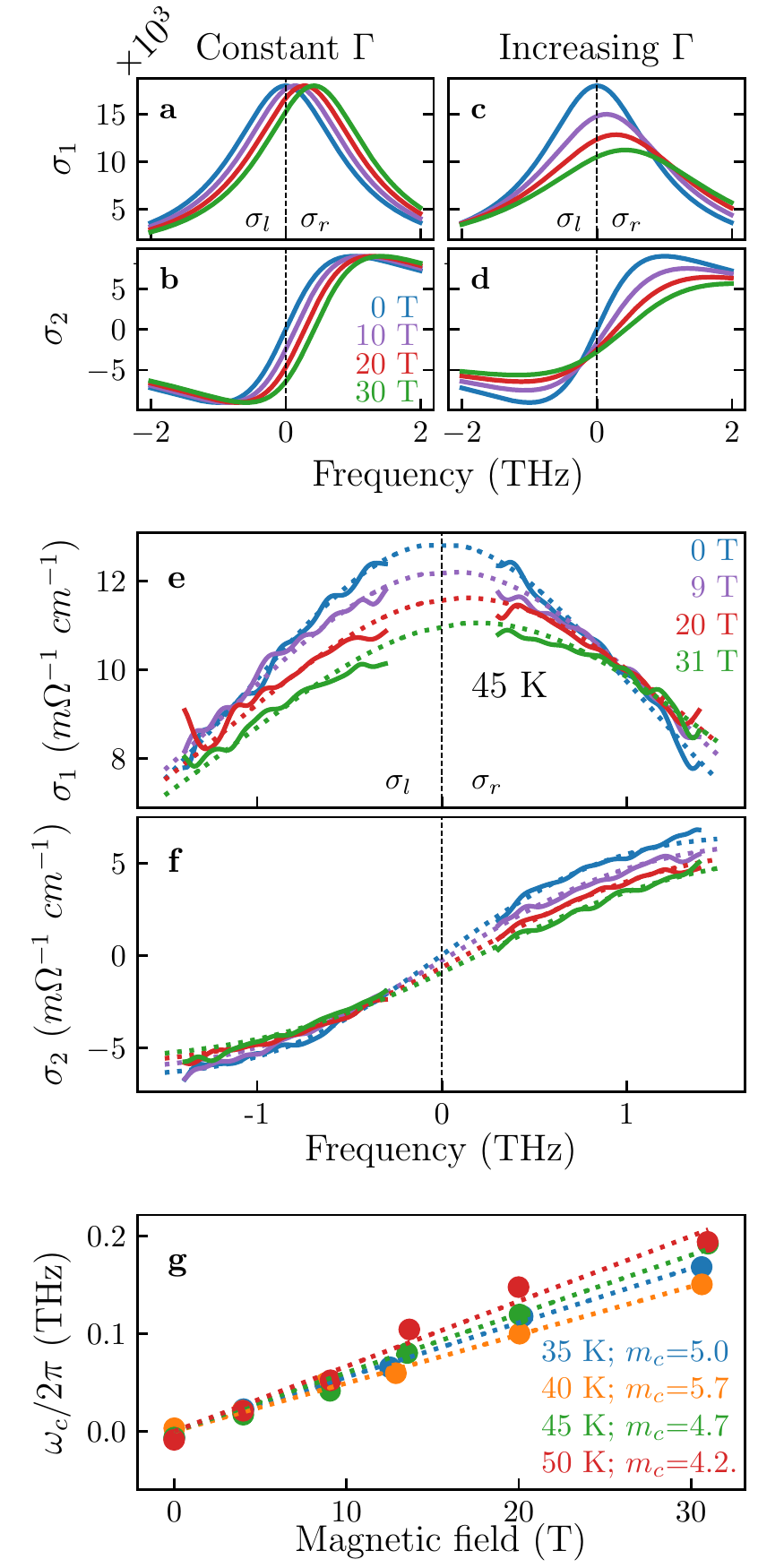}
\caption{\label{Fig3} a,b) Example of circularly-polarized complex conductivity for a model system exhibiting simple CR (using $m_{\rm c}=2~m_{\rm e}$, $\Gamma$=1~THz, and a fixed spectral weight).  Negative and positive frequencies correspond to $\sigma_l(\omega)$ and $\sigma_r(\omega)$, respectively. Shown in this way, the Drude peak appears as a Lorentzian oscillator centered about $\omega_{\rm c} = eB/m_c$. c,d) Same, except now $\Gamma$ increases linearly from 1.0 to 1.6~THz as $B \rightarrow 30$~T. The Lorentzian conductivity peak now also broadens and diminishes in amplitude with increasing $B$. e,f) Experimentally measured optical conductivity from LSCO up to 31~T, at 45~K. A clear cyclotron shift is observed, along with an amplitude reduction and broadening consistent with increasing $\Gamma$.  Dashed lines show fits using a cyclotron-active two-fluid model. g) The extracted values of $\omega_{\rm c}$. Linear fits to $\omega_{\rm c} (B)$ reveal the cyclotron mass $m_c$.}
\end{figure}

A cyclotron shift of the optical conductivity in a magnetic field is most easily observed by plotting $\sigma_r(\omega)$ and $\sigma_l(\omega)$, the cyclotron-active and -inactive modes, on the same axis using positive and negative values of frequency, respectively. In the simplest model of CR, the real ($\sigma_1$) and imaginary ($\sigma_2$) conductivities then take the form illustrated in Figs. \ref{Fig3}a,b. At $B$=0 the Drude conductivity peak appears as a  single Lorentzian oscillator centered at zero frequency with width $\Gamma$. Applied fields $B$ simply shift the Drude peak from zero to $\omega_{\rm c}$, as depicted \cite{palikInfraredMicrowaveMagnetoplasma1970, chengMagnetoterahertzResponseFaraday2019}. However, if $\Gamma$ also increases with $B$, then $\sigma_1$ and $\sigma_2$ are additionally broadened and reduced in amplitude as shown in Figs. \ref{Fig3}c,d. 

Figures 3e and 3f show the experimentally-measured complex conductivity, at 45~K (above $T_{\rm c}$), at selected magnetic fields up to 31~T. As field is increased, the data exhibit a small but clear shift of the peak position in the real part of the conductivity, accompanied by a reduction of the peak amplitude, similar to the model depicted in Figs. \ref{Fig3}c,d. To quantify the observed behavior, we fit these spectra using a modified two-fluid model that includes a cyclotron shift $\omega_{\rm c}$:
\begin{equation}
\label{Eqn4}
\sigma_{r,l}(\omega) = i\epsilon_0
\left( \frac{\omega^2_{\rm p,n}}{\omega - \omega_{\rm c} +i \Gamma } + \frac{\omega_{\rm p,s}^2}{\omega} \right). 
\end{equation}
Note that the sign of $\omega_{\rm c}$ depends on the sign of the charge, and that $\sigma_r(\omega)$ and $\sigma_l(\omega)$ are defined for positive and negative frequencies, as described above. We include the small $\omega^2_{\rm p,s}$ term to account for inertial effects of the fluctuating superconductivity that can exist in these samples slightly above $T_c$ \cite{bilbroTemporalCorrelationsSuperconductivity2011a}. Best fits are shown by the dashed lines (fits to all data sets are shown in the Supplementary Information) and the measured values of $\omega_{\rm c}$ are plotted in Fig. \ref{Fig3}g. A linear fit of $\omega_{\rm c}$ to $B$ indicates a relatively heavy carrier mass $m_c \approx 4.7~m_{\rm e}$ at 45~K. To confirm this result, Fig. 3g also shows results of additional measurements performed at nearby temperatures, each yielding similar mass values, which together imply $m_{\rm c}=4.9\pm 0.8$ $m_{\rm e}$. Note that it was essential here to measure the complex conductivity in the circular basis to resolve the small frequency shift of the conductivity peak.  Because of the peak's comparatively large width, measurements sensitive only to $\sigma_{xx}$ would have shown primarily a drop in the peak amplitude and a broadening, but not a shift.

Finally, the extracted spectral weights of the normal and superconducting carriers are shown in Fig. 4a and the scattering rates in Fig. 4b.  With increasing $B$, $\omega_{\rm p,n}^2$ increases and saturates near the same value for all measured temperatures, while $\omega_{\rm p,s}^2$ is suppressed to zero, as expected for field-induced suppression of superconductivity. Concomitantly, the extracted scattering rate $\Gamma$ (Fig. 4b) is found to increase with $B$, by an amount that is temperature-dependent. The field dependence of $\Gamma$ can also be directly seen in the conductivity spectra of Fig. 3e, via the amplitude reduction and broadening of $\sigma_1(\omega)$ as $B$ increases.  

\begin{figure}
\centering
\includegraphics[width=.9\columnwidth]{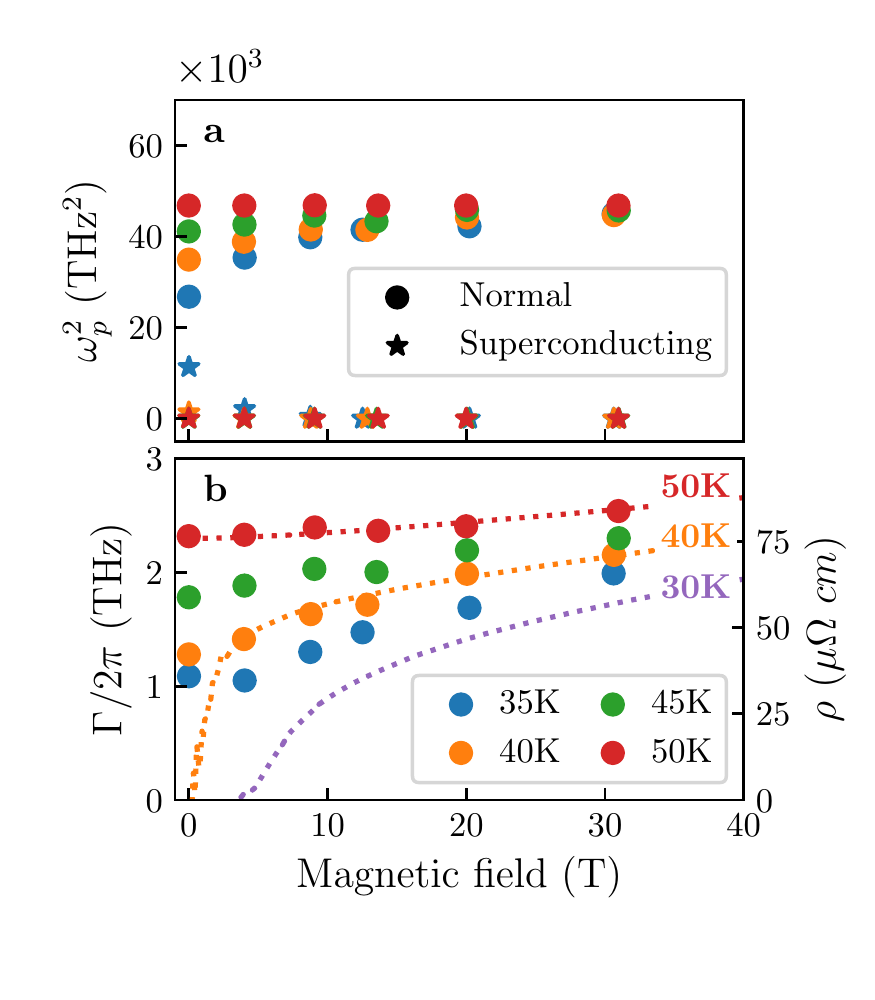}
\caption{\label{Fig4} a) Magnetic field dependence of the spectral weights $\omega_{\rm p,n}^2$ and $\omega_{\rm p,s}^2$ at all measured temperatures. $\omega_{\rm p,s}^2$  decreases as superconductivity is suppressed by $B$, while $\omega_{\rm p,n}^2$ increases. b) The scattering rate $\Gamma$ increases with $B$ at all measured temperatures. Also shown is $\rho$ (dotted lines, right axis), the resistivity  measured in a similarly-doped LSCO thin film, at 50, 40, and 30~K (from \cite{giraldo-galloScaleinvariantMagnetoresistanceCuprate2018}).}
\end{figure}

These results represent the first direct measurement of CR in the La$_{2-x}$Sr$_{x}$CuO$_4$ system. Comparing this hole effective mass (4.9~$m_e$) to values obtained from quantum oscillation studies in other HTSC materials, it is larger than the mass inferred in optimally-doped YBCO ($3.6~m_e$) \cite{ramshawQuasiparticleMassEnhancement2015}, and is closer to values obtained in highly overdoped Tl$_2$Ba$_2$CuO$_{6+\delta}$ (4.1 $m_{\rm e}$) \cite{vignolleQuantumOscillationsOverdoped2008}.  We note that hole masses previously estimated in LSCO from studies that combined zero-field optics and Hall effect data ($ \approx$4~$m_e$) \cite{padillaConstantEffectiveMass2005} are comparable to our measured $m_c$.  Moreover, a comparison of the observed cyclotron effective mass with photoemission and electrical transport experiments \cite{horioThreeDimensionalFermiSurface2018, fangFermiSurfaceTransformation2020} can be made via a tight-binding parameterization of the band structure \cite{MarkiewiczTightbinding}. At a Fermi energy corresponding to $16\%$ hole doping, such comparisons predict a cyclotron mass of 4.8~$m_{\rm e}$ (see also Supplemental Information).  It is important to note that this treatment potentially misses renormalization occuring at very low energies close to $E_{\rm F}$, and near renormalization ``hot-spots'' at other regions in Brillouin zone.  Furthermore, the observed value of the cyclotron effective mass at $x = 0.16$ is in broad agreement with the reported Sommerfeld coefficient of electronic heat capacity \cite{Momono1994, Michon2019}. Given that these different experimental determinations of effective mass are in principle subject to differing degrees of mass renormalization, the extent to which the masses measured through CR agree with those extracted from other techniques -- particularly over a wide range of doping levels -- should be an important topic for future inquiry. 

In addition to the carrier masses, other important parameters extracted from these THz studies (such as the scattering rate $\Gamma$) can be compared to values obtained by complementary methods, such as electrical transport. Figure 4b shows a comparison between $\Gamma$ extracted from our THz studies, and the resistivity $\rho (B)$ (dashed lines) reported in a similar LSCO film at similar temperatures \cite{giraldo-galloScaleinvariantMagnetoresistanceCuprate2018}. Both $\Gamma (B)$ and $\rho (B)$ follow the same general trends, implying that these quantities are closely related. With further analysis, this observation could help distinguish between competing models for the distinctive normal-state magnetoresistance of cuprates \cite{singletonSimpleTransportModel2018}.

We emphasize that LSCO can be tuned across its entire doping range, in contrast to other hole-doped cuprates in which quantum oscillations have been measured (YBCO, HgBa$_2$CuO$_{4+\delta}$, and Tl$_2$Ba$_2$CuO$_{6+\delta}$).  Accordingly, the successful measurement of $\omega_{\rm c}$ demonstrated in this work paves the way for future studies aimed at characterizing the full doping dependence of $m_{\rm c}$. We note that both the van Hove singularity in the band structure \cite{Yoshida2009, Chang2008} at a doping $p \approx 0.19 \approx p*$ (the pseudogap critical point in LSCO \cite{Cooper2009}), and the reported divergence of the electronic heat capacity in the same doping region \cite{Michon2019}, suggest a rapidly-varying effective mass in this doping range. Thus, CR measurements are well positioned to shed significant light on the role of single-particle \textit{vs.} many-body renormalization effects in the cuprates.  Combined with direct measurement of $\Gamma$ at temperatures above and below $T_{\rm c}$ that are enabled by TDTS, such results have the potential to bring new insights into the role of Coulomb interactions in HTSC, specifically ``strange metal'' phenomena such as a $T$-linear resistivity at low temperatures, and the role of quantum criticality in shaping the superconducting dome.

Work at the National High Magnetic Field Laboratory was supported by National Science Foundation (NSF) DMR-1644779, the State of Florida, and the U.S. Department of Energy (DOE). AL and NPA were supported by the Quantum Materials program at the Canadian Institute for Advanced Research and NSF DMR 1905519. Work at Brookhaven National Laboratory was supported by the DOE, Basic Energy Sciences, Materials Sciences and Engineering Division. X. H. is supported by the Gordon and Betty Moore Foundation's EPiQS Initiative through grant GBMF9074. We gratefully acknowledge G. Granroth for providing the CuAg wire used in our pulsed magnet, and S. Chakravarty, J. Chang, S. Chen, M. Horio,  S. Kivelson, B. Ramshaw, A. Shekter, Z. X. Shen, and L. Taillefer for helpful conversations.

%\bibliography{MyLibrary2}{}
%\bibliographystyle{plain}

\end{document}